\documentclass[12pt,twoside,english]{article}
\usepackage[T1]{fontenc}
\usepackage[latin9]{inputenc}
\usepackage{geometry}
\usepackage{color}
\usepackage[greek,british]{babel}
\usepackage{amsmath}
\usepackage{amssymb}
\usepackage[unicode=true,
 bookmarks=true,bookmarksnumbered=true,bookmarksopen=true,bookmarksopenlevel=1,
 breaklinks=false,pdfborder={0 0 1},backref=false,colorlinks=true]
 {hyperref}
\hypersetup{pdftitle={The LyX Tutorial},
 pdfauthor={LyX Team},
 pdfsubject={LyX-documentation Tutorial},
 pdfkeywords={LyX, documentation},
 linkcolor=black, citecolor=black, urlcolor=blue, filecolor=blue,pdfpagelayout=OneColumn, pdfnewwindow=true, pdfstartview=XYZ, plainpages=false}
\usepackage{breakurl}

\makeatletter
\usepackage{ifpdf} 
\ifpdf 

 \IfFileExists{lmodern.sty}{\usepackage{lmodern}}{}

\fi 

\let\myTOC\tableofcontents
\renewcommand\tableofcontents{%
  \frontmatter
  \pdfbookmark[1]{\contentsname}{}
  \myTOC
  \mainmatter }

\def\LyX{\texorpdfstring{%
  L\kern-.1667em\lower.25em\hbox{Y}\kern-.125emX\@}
  {LyX}}

\makeatother

\begin{document}

\title{The Netrinos of the Neighboring Brane}

\author{{\Large A. Nicolaidis}\medskip \\Theoretical Physics Department\\Aristotle
University of Thessaloniki\\54124 Thessaloniki, Greece\\nicolaid@auth.gr}
\maketitle
\begin{abstract}
The phenomenon of neutrino oscillations is studied usually as a mixing
between the flavor neutrinos and the neutrinos having a definite mass.
The mixing angles and the mass eigenvalues are treated independently
in order to accommodate the experimental data. We suggest that neutrino
oscillations are connected to the structure of spacetime. We expand
on a recently proposed model, where two \textquotedblleft{}mirror\textquotedblright{}
branes coexist. One brane hosts left-handed particles (our brane),
while the other brane hosts right-handed particles. Majorana-type
couplings mixes neutrinos in an individual brane, while Dirac-type
couplings mixes neutrinos across the brares. We first focus our attention
in a single brane. The mass matrix, determined by the Majorana mass,
leads to mass eigenstates and further to mixing angles identical to
the mixing angles proposed by the tri-bimaximal mixing. When we include
the Dirac-type coupling, connecting the two branes, we obtain a definite
prediction for the transition to a sterile neutrino (right-handed
neutrino). With $m_{L}$ ($m_{R}$) the Majorana mass for the left
(right) brane, we are able to explain the solar and the atmospheric
neutrino data with $m_{L}=2m_{R}$ and $m_{R}=10^{-2}$ eV.
\end{abstract}
\newpage

\qquad{}Neutrinos, more than 80 years from their inception, remain
enigmatic. A number of experiments have helped us to determine their
mixing angles and the scale of their masses \cite{key-1}. Yet, we
are lacking a satisfactory explanation for the nature of neutrinos,
their number and the actual values of the parameters involved. In
the present paper we attempt to connect the neutrino issues with the
fundamental problem of theoretical physics, namely the structure of
spacetime. 

\qquad{}The entire universe (matter and radiation, stars, galaxies)
is under a continuous evolution. Should we evolve also our own notions
of space and time, should we look for a dynamical emergence of spacetime?
Recently, by using the Cartan-Penrose connection of spinors to geometry,
we explored the geometrical structures consistent with the quantum
entanglement of two spinors \cite{key-2}. Let us remind the thrust
of the argument.

\qquad{}Relational logic, or its equivalent formulation as category
theory, has been presented as the common foundation of quantum mechanics
and string theory \cite{key-3}. With relation (or a morphism) represented
by a spinor \cite{key-3,key-4}, we adopted the Cartan method of using
spinors to obtain linear representations of geometries \cite{key-5}.
A single spinor gives rise to the Riemann-Bloch sphere, which is topologically
equivalent to the null cone of Minkowski spacetime \cite{key-6}.
It is quite natural then to wonder what kind of geometry we obtain,
when we entangle two spinors.

\qquad{}There are two ways to couple two spinors. The first recipe
is coming from Majorana \cite{key-7}. Given a left-handed spinor
$\left|\psi_{L}\right\rangle $, we may construct a right-handed spinor
$\left|\chi_{R}\right\rangle $ by 
\begin{equation}
\left|\chi_{R}\right\rangle =\sigma_{2}\left|\psi_{L}\right\rangle ^{*}
\end{equation}

Starting with two independent left-handed Weyl spinors, we may induce
a coupling between them by establishing a four-component Majorana
spinor
\begin{equation}
\left|\Psi_{M}\right\rangle =\left(\begin{array}{c}
\left|\chi_{L}\right\rangle \\
\sigma_{2}\left|\psi_{L}\right\rangle ^{*}
\end{array}\right)
\end{equation}
Defining $X_{i}=\left\langle \Psi_{M}\right.\left|\gamma_{i}\right|\left.\Psi_{M}\right\rangle $
$\left(i=0,1,2,3\right)$ we find that $X_{i}$ is not a null vector
\cite{key-2}
\begin{equation}
X_{1}^{2}+X_{2}^{2}+X_{3}^{2}-X_{0}^{2}=M_{M}^{2}
\end{equation}
with
\begin{eqnarray*}
X_{4} & = & i\left\langle \Psi_{M}\right|\left.\Psi_{M}\right\rangle \\
X_{5} & = & \left\langle \Psi_{M}\right.\left|\gamma_{5}\right|\left.\Psi_{M}\right\rangle 
\end{eqnarray*}
\begin{equation}
M_{M}^{2}=-\left(X_{4}^{2}+X_{5}^{2}\right)
\end{equation}
Thus among two left-handed Weyl spinors (or two right-handed Weyl
spinors), the Majorana's coupling induces a mass term.

\qquad{}The Dirac coupling involves a left-handed Weyl spinor and
a right-handed Weyl spinor. Writing
\begin{equation}
\left|\Psi_{D}\right\rangle =\left(\begin{array}{c}
\left|\chi_{L}\right\rangle \\
\left|\psi_{R}\right\rangle 
\end{array}\right)
\end{equation}
we obtain 
\begin{equation}
X_{1}^{2}+X_{2}^{2}+X_{3}^{2}-X_{0}^{2}=-M_{D}^{2}\label{eq:6}
\end{equation}
with
\begin{equation}
M_{D}^{2}=\left(X_{4}^{2}+X_{5}^{2}\right)
\end{equation}

Let us define $T=X_{0}$, $t=M_{D}$. The Dirac entanglement, equ.(\ref{eq:6}),
takes the form of a space-like hyperboloid
\begin{equation}
T^{2}-\sum_{i=1}^{3}X_{i}^{2}=t^{2}\label{eq:8}
\end{equation}
A comparison with the null cone geometry, indicates that quantum entanglement,
specified and quantified by t, generates an extra dimension. The distance
along this extra dimension indicates how far we are from the null
cone. Furthermore our space-time acquires a double-sheet structure,
reminding the ekpyrotic model where two branes coexist \cite{34c,35c,36c}.
There is though a distinct difference. In our model, by construction,
one brane hosts left-handed particles (our brane), while the other
brane hosts right-handed particles.

\qquad{}The conventional way to restore left-right symmetry is to
introduce an extra $SU(2)_{R}$ gauge group in the energy desert above
the scale of the standard $SU(2)_{L}$ interactions. The right-handed
gauge bosons are more massive compared to the left-handed gauge bosons,
leading to parity violation at low energies \cite{37c,38c}. Within
our approach the left-right symmetry is achieved with the extra dimension
hosting two ``mirror'' branes, a left-handed brane and a right-handed
brane. The most prominent candidate for mediation between the two
branes is the neutrino particle. The left-handed neutrino, an essential
ingredient of the standard model, resides in our brane, while its
counterpart, the right-handed neutrino, resides in the other brane.
Within our approach neutrino oscillations acquire a novel character.
Majorana-type coupling mixes the left-handed flavor neutrinos residing
in our brane, as well as the right-handed neutrinos residing in the
other brane. Dirac-type coupling connects the left-handed neutrinos
of our brane to the right-handed neutrinos of the other brane. From
our point of view, right-handed neutrinos appear as sterile neutrinos,
and the transition flavor neutrino - sterile neutrino - flavor neutrino
amounts to a swapping between the two branes. Let us study first the
mixing among the left-handed neutrinos, or focus our attention into
our brane.

\subparagraph*{i) single brane}

\qquad{}We assume a ``democratic principle'' attributing the same
value to all Majorana mass couplings. Then the mass matrix for the
left-handed neutrinos will take the form
\begin{equation}
M=\left(\begin{array}{ccc}
0 & m & m\\
m & 0 & m\\
m & m & 0
\end{array}\right)\label{eq:9}
\end{equation}
The eigenvalues of $M$, involving a double root, are
\begin{equation}
\lambda_{1}=\lambda_{3}=-m\qquad\lambda_{2}=2m
\end{equation}
The corresponding eigenvectors are
\begin{eqnarray}
N_{1}^{T} & = & \frac{1}{\sqrt{6}}\left(\begin{array}{ccc}
2, & -1, & -1\end{array}\right)\nonumber \\
N_{2}^{T} & = & \frac{1}{\sqrt{3}}\left(\begin{array}{ccc}
1, & 1, & 1\end{array}\right)\label{eq:11}\\
N_{3}^{T} & = & \frac{1}{\sqrt{2}}\left(\begin{array}{ccc}
0, & 1, & -1\end{array}\right)\nonumber 
\end{eqnarray}
Expressing the flavor left-handed neutrinos in terms of the mass eigenstates
we write
\begin{equation}
\left|\nu_{f_{i}}\right\rangle =\sum_{j}c_{ij}\left|N_{j}\right\rangle 
\end{equation}
with $\nu_{f_{1}},\nu_{f_{2}},\nu_{f_{3}}$ denoting respectively
$\nu_{e_{L}},\nu_{\mu_{L}},\nu_{\tau_{L}}$.

\qquad{}Defining $\left(U\right)_{ij}=c_{ij}$ we find that the mixing
matrix $U$ is
\begin{equation}
U=\left(\begin{array}{ccc}
\sqrt{\frac{2}{3}} & \frac{1}{\sqrt{3}} & 0\\
-\frac{1}{\sqrt{6}} & \frac{1}{\sqrt{3}} & \frac{1}{\sqrt{2}}\\
-\frac{1}{\sqrt{6}} & \frac{1}{\sqrt{3}} & -\frac{1}{\sqrt{2}}
\end{array}\right)\label{eq:13}
\end{equation}

This type of mixing defines the celebrated tri-bimaximal mixing (TB
mixing) {[}13-16{]}. We notice that the TB mixing has been proposed
in order to accomodate the experimental data, while in our case emerges
as the outcome of a Majorana-type coupling among the left-handed neutrinos.

\qquad{}Let us consider an initial flavor $\nu_{e_{L}}$ beam. The
transitions to other flavors are given by 
\begin{equation}
P\left(\nu_{e}\rightarrow\nu_{\mu}\right)=P\left(\nu_{e}\rightarrow\nu_{\tau}\right)=\frac{4}{9}sin^{2}\left(\frac{3}{4}\frac{m^{2}}{\hbar E}t\right)
\end{equation}
Also
\begin{equation}
P\left(\nu_{e}\rightarrow\nu_{e}\right)=\frac{1}{9}\left[9-\frac{8}{9}sin^{2}\left(\frac{3}{4}\frac{m^{2}}{\hbar E}t\right)\right]
\end{equation}
The oscillations depend upon a single mass scale and clearly cannot
reproduce the available data. The introduction of the right-handed
brane allows us to have access to two more scales, the Majorana mass
coupling in the right-handed brane and the Dirac mass coupling among
the branes. We move then to the case of the two ``mirror'' branes.

\subparagraph{ii) mirror branes}

\qquad{}On general grounds we expect the Majorana mass coupling in
the right-handed brane to be of the same order of magnitude with the
corresponding parameter in the left-handed brane. For general purposes
we denote them by $m_{L}$, $m_{R}$, with the obvious correspondence.
Each single left-handed neutrino, residing in our brane, is connected
to all the right-handed neutrinos, residing in the other brane, by
the same universal Dirac mass coupling $\mu$. Then the mass matrix
involving the 6 neutrino states (3 left-handed plus 3 right-handed)
will have the form
\begin{equation}
\mathcal{M}=\left(\begin{array}{cc}
M_{L} & M_{+}\\
M_{+} & M_{R}
\end{array}\right).
\end{equation}
$M_{L}\left(M_{R}\right)$ is a mass matrix identical to $M$, equ.
(\ref{eq:9}), with $m$ replaced by $m_{L}\left(m_{R}\right)$. $M_{+}$
involves the mass terms connecting the two branes and is given by
\begin{equation}
M_{+}=\left(\begin{array}{ccc}
\mu & \mu & \mu\\
\mu & \mu & \mu\\
\mu & \mu & \mu
\end{array}\right)
\end{equation}
The eigenvalues, involving two double roots, are 
\begin{equation}
\begin{aligned}\lambda_{1}= & \lambda_{3}=-m_{L}\\
\lambda_{4}= & \lambda_{6}=-m_{R}\\
\lambda_{2}= & \left(m_{L}+m_{R}\right)+\left[\left(m_{L}-m_{R}\right)^{2}+9\mu^{2}\right]^{\frac{1}{2}}\\
\lambda_{5}= & \left(m_{L}+m_{R}\right)-\left[\left(m_{L}-m_{R}\right)^{2}+9\mu^{2}\right]^{\frac{1}{2}}
\end{aligned}
\end{equation}
Let us define
\begin{equation}
\begin{aligned}d= & \left[\left(m_{L}-m_{R}\right)^{2}+9\mu^{2}\right]^{\frac{1}{2}}\\
\delta_{\pm}= & d\pm\left(m_{L}-m_{R}\right)
\end{aligned}
\end{equation}
\begin{equation}
\cos\phi=\left(\frac{\delta_{+}}{2d}\right)^{\frac{1}{2}}\qquad\sin\phi=\left(\frac{\delta_{-}}{2d}\right)^{\frac{1}{2}}
\end{equation}
The corresponding eigenvectors are 
\begin{equation}
\begin{aligned}N_{1}^{T}= & \frac{1}{\sqrt{6}}\left(\begin{array}{cccccc}
2, & -1, & -1, & 0, & 0, & 0\end{array}\right)\\
N_{2}^{T}= & \frac{1}{\sqrt{3}}\left(\begin{array}{cccccc}
\cos\phi, & \cos\phi, & \cos\phi, & \sin\phi, & \sin\phi, & \sin\phi\end{array}\right)\\
N_{3}^{T}= & \frac{1}{\sqrt{2}}\left(\begin{array}{cccccc}
0, & 1, & -1, & 0, & 0, & 0\end{array}\right)\\
N_{4}^{T}= & \frac{1}{\sqrt{6}}\left(\begin{array}{cccccc}
0, & 0, & 0, & 1, & -2, & 1\end{array}\right)\\
N_{5}^{T}= & \frac{1}{\sqrt{3}}\left(\begin{array}{cccccc}
\sin\phi, & \sin\phi, & \sin\phi, & -\cos\phi, & -\cos\phi, & -\cos\phi\end{array}\right)\\
N_{6}^{T}= & \frac{1}{\sqrt{2}}\left(\begin{array}{cccccc}
0, & 0, & 0, & -1, & 0, & 1\end{array}\right)
\end{aligned}
\end{equation}

The similarities and the differences with the case of a single brane,
equ. (\ref{eq:11}), are apparent. $N_{1},N_{3}\left(N_{4},N_{6}\right)$
involve mixing within the individual left (right) brane. $N_{2}$
and $N_{5}$ connect the two branes. For $m_{L}=m_{R}$, $\phi=\frac{\pi}{4}$
and the two branes are equally present in the mixing phenomenon. For
small Dirac coupling compared to the Majorana couplings we obtain
$\phi\simeq0$.

\qquad{}The mixing matrix connecting the flavor eigenstates (left-handed
and right-handed) to the six eigenvectors takes the form
\begin{equation}
U=\left(\begin{array}{cccccc}
\sqrt{\frac{2}{3}} & \frac{1}{\sqrt{3}}\cos\phi & 0 & 0 & \frac{1}{\sqrt{3}}\sin\phi & 0\\
-\frac{1}{\sqrt{6}} & \frac{1}{\sqrt{3}}\cos\phi & \frac{1}{\sqrt{2}} & 0 & \frac{1}{\sqrt{3}}\sin\phi & 0\\
-\frac{1}{\sqrt{6}} & \frac{1}{\sqrt{3}}\cos\phi & -\frac{1}{\sqrt{2}} & 0 & \frac{1}{\sqrt{3}}\sin\phi & 0\\
0 & \frac{1}{\sqrt{3}}\sin\phi & 0 & \frac{1}{\sqrt{6}} & -\frac{1}{\sqrt{3}}\cos\phi & -\frac{1}{\sqrt{2}}\\
0 & \frac{1}{\sqrt{3}}\sin\phi & 0 & -\sqrt{\frac{2}{3}} & -\frac{1}{\sqrt{3}}\cos\phi & 0\\
0 & \frac{1}{\sqrt{3}}\sin\phi & 0 & \frac{1}{\sqrt{6}} & -\frac{1}{\sqrt{3}}\cos\phi & \frac{1}{\sqrt{2}}
\end{array}\right)
\end{equation}
Again for $\phi=0$ the upper left part of the matrix gives the previous
result, equ. (\ref{eq:13}), for the single brane.

\qquad{}Imagine that at $t=0$ we start with a pure $\nu_{e_{L}}$
beam. The probability to find later another flavor is given by
\begin{equation}
\begin{aligned}P\left(\nu_{e_{L}}\rightarrow\nu_{\mu_{L}}\right)=P\left(\nu_{e_{L}}\rightarrow\nu_{\tau_{L}}\right)= & \frac{1}{9}\left\{ 1+\cos^{4}\phi+\sin^{4}\phi+\right.\\
 & 2\left[\cos^{2}\phi\sin^{2}\phi\cos\frac{\left(\omega_{+}-\omega_{-}\right)t}{2\hbar E}\right.\\
 & \left.\left.-\cos^{2}\phi\cos\frac{\omega_{+}t}{2\hbar E}-\sin^{2}\phi\cos\frac{\omega_{-}t}{2\hbar E}\right]\right\} 
\end{aligned}
\end{equation}
where
\begin{equation}
\begin{aligned}\omega_{+}= & m_{L}^{2}+2m_{R}^{2}+9\mu^{2}+2d\left(m_{R}+m_{L}\right)\\
\omega_{-}= & m_{L}^{2}+2m_{R}^{2}+9\mu^{2}-2d\left(m_{R}+m_{L}\right)\\
\omega_{+}-\omega_{-}= & 4d\left(m_{R}+m_{L}\right)
\end{aligned}
\end{equation}
The transition to a generic sterile neutrino (an incoherent sum of
all right-handed neutrinos) is given by 
\begin{equation}
P\left(\nu_{e}\rightarrow\nu_{s}\right)=\frac{1}{9}\sin2\phi\sin^{2}\left(\frac{1}{4\hbar E}\left(\omega_{+}-\omega_{-}\right)t\right)\label{eq:25}
\end{equation}
Notice that for the transition of the $\nu_{\mu_{L}}$ we find 
\begin{equation}
P\left(\nu_{\mu_{L}}\rightarrow\nu_{e_{L}}\right)=P\left(\nu_{\mu_{L}}\rightarrow\nu_{\tau_{L}}\right)=P\left(\nu_{e_{L}}\rightarrow\nu_{\mu_{L}}\right)
\end{equation}

\qquad{}We may recall the neutrino oscillation data \cite{24c}.
Solar and atmospheric neutrino oscillations define two distinct mass
scales
\begin{equation}
\Delta m_{s}^{2}\simeq5*10^{-5}eV^{2}\qquad\Delta m_{a}^{2}\simeq2*10^{-3}eV^{2}
\end{equation}
A neutrino oscillation experiment defines a specific value for the
parameter $\tfrac{t}{E}$ (the distance traveled by the neutrino over
its energy). Large values of $\tfrac{t}{E}$ allow to explore small
values of $\Delta m^{2}$, or correspondingly small $\omega$. Solar
neutrinos correspond to low energy neutrinos covering huge distance,
therefore their oscillation is determined by $\omega_{-}$. Atmospheric
neutrinos involve higher energies and smaller distances and their
oscillation is controlled by $\omega_{+}$. Accordingly we assign
\begin{equation}
\begin{aligned}\omega_{-}\simeq & \Delta m_{s}^{2}\\
\omega_{+}\simeq & \Delta m_{a}^{2}
\end{aligned}
\label{eq:28}
\end{equation}
There is a conflicting evidence for the existence of a sterile neutrino
\cite{key-18}. At any rate the amplitude for a transition to a sterile
neutrino is expected to be small and correspondingly $\sin\phi$,
see equ. (\ref{eq:25}), and the Dirac coupling $\mu$ are small.
Adopting the hierarchy $\left(m_{L}-m_{R}\right)>\mu$ we find that
the values 
\begin{equation}
m_{L}\simeq2m_{R}\qquad m_{R}\simeq10^{-2}eV
\end{equation}
reproduce the observed scales, equ. (\ref{eq:28}). The precise smallness
of $\mu$ will fix the magnitude of $\sin\phi$ and therefore the
probability to a sterile neutrino oscillation. Notice however that
within our scheme the mass scale for the transition to a sterile neutrino
is at a sub-eV scale ($3*10^{-2}$ eV), rather far from the value
suggested by the LSND experiment.

\qquad{}The conventional approach to the phenomenon of neutrino oscillations
is to consider it as a manifestation of a mixing between the flavor
eigenstates and the mass eigenstates. The mixing angles and the masses
of the mass eigenstates are treated independently and are determined
largely by the experimental data. There is also an effort to accommodate
the available data by making appeal to discrete groups \cite{key-19}.
We offer an alternative approach, by proposing that neutrino oscillations
are connected to the structure of space-time. Space-time hosts two
branes, one brane where the left-handed particles reside (our brane)
and another brane where the right-handed particles reside. The long
sought left-right symmetry is achieved through the geometry of space-time.
Majorana-type couplings connect the neutrinos living in an individual
brane, while Dirac-type couplings connect neutrinos across the branes.
We managed to treat at the same time both the masses involved and
the mixing angles, by making appeal to first principles. Is this success
fortuitous? We may argue that it is a sign for the existence of \textquotedblleft{}mirror\textquotedblright{}
branes. But clearly further indications are needed.

\qquad{}Finally we would like to remind the experimental evidence
for a small non-vanishing value for the matrix element $c_{13}$ {[}20-23{]}.
It seems that this small value indicates a hidden substructure and
work along this line is in progress.

\section*{Acknowledgments}

A useful discussion with Dr. Ion Siotis is gratefully acknowledged.
The present work is supported by the Templeton Foundation.

\end{document}